\documentclass[letterpaper,journal,twoside]{IEEEtran}
\usepackage{amsmath,amsfonts}
\usepackage[ruled,vlined]{algorithm2e}
\usepackage{subcaption}  
\usepackage{booktabs}
\usepackage{array}
\usepackage[caption=false,font=normalsize,labelfont=sf,textfont=sf]{subfig}
\usepackage{textcomp}
\usepackage{stfloats}
\usepackage{url}
\usepackage{verbatim}
\usepackage{graphicx}
\usepackage{cite}
\begin{document}
\title{
Toward Robust Single-Photon Perception for Robots: A Condition-Aware Active Learning Approach
}

\author{
Zili Zhang$^{1,\dagger}$, Ziting Wen$^{2,\dagger}$, Yiheng Qiang$^{1}$,
Hongzhou Dong$^{1}$, Wenle Dong$^{1}$,\\
Xinyang Li$^{1}$, Kemi Ding$^{2}$, Xiaofan Wang$^{1}$,
Guodong Shi$^{3}$, and Xiaoqiang Ren$^{1,4,*}$%
\thanks{Manuscript received: December 28, 2025; Revised: June 29, 2026; Accepted: August 2, 2026.}
\thanks{This paper was recommended for publication by Editor Abhinav Valada upon evaluation of the Associate Editor and Reviewers comments.}
\thanks{$^{1}$Zili Zhang, Yiheng Qiang, Hongzhou Dong, Wenle Dong,
Xinyang Li, Xiaofan Wang, and Xiaoqiang Ren are with the
School of Mechatronic Engineering and Automation,
Shanghai University, Shanghai 200444, China.}%
\thanks{$^{2}$Ziting Wen and Kemi Ding are with the
School of Automation and Intelligent Manufacturing,
Southern University of Science and Technology,
Guangdong 518055, China.}%
\thanks{$^{3}$Guodong Shi is with the Australian Centre for Robotics,
School of Aerospace, Mechanical and Mechatronic Engineering,
University of Sydney, Sydney, NSW 2006, Australia.}%
\thanks{$^{4}$Xiaoqiang Ren is also with the School of Future Technology
and the National Key Laboratory of Maritime Unmanned Swarm Intelligence,
Shanghai University, Shanghai 200444, China.}%
\thanks{$^{\dagger}$Zili Zhang and Ziting Wen contributed equally to this work.}%
\thanks{$^{*}$Corresponding author: Xiaoqiang Ren
(\texttt{xqren@shu.edu.cn}).}%
\thanks{Digital Object Identifier (DOI): see top of this page.}
}

\markboth
{IEEE ROBOTICS AND AUTOMATION LETTERS. PREPRINT VERSION. AUGUST, 2026}%
{Zhang \MakeLowercase{\textit{et al.}}: Toward Robust Single-Photon Perception for Robots: A Condition-Aware Active Learning Approach}

\maketitle

\begin{abstract} 
LiDAR-based perception plays a fundamental role in modern robotic systems for environment understanding and navigation. Single-photon LiDAR (SPL) extends conventional LiDAR by enabling photon-efficient 3D sensing under challenging conditions such as long-range operation, low-albedo targets, and limited signal returns. However, developing SPL perception models for real-world robotic applications remains difficult because annotated SPL data are costly to obtain and model performance can vary substantially across imaging conditions. In this paper, we present the first active learning framework tailored to the SPL sensing modality rather than a specific downstream task. Our method introduces a physics-grounded, imaging condition-aware sampling strategy that uses synthetic SPL variants to characterize how candidate samples respond to changes in sensing conditions. By jointly modeling prediction uncertainty, sample diversity, and sensitivity to photon-level imaging variations, the proposed approach prioritizes samples that are informative for improving labeling efficiency and robustness. Extensive experiments on synthetic and real-world SPL datasets demonstrate that our method substantially reduces annotation requirements while maintaining strong performance across image-level and dense prediction settings. On synthetic data, our approach achieves 97\% classification accuracy using only 1.5\% labeled samples. On real-world data, it maintains 90\% accuracy with 8.2\% labeled samples, outperforming the strongest baseline by 6\%. Segmentation results further show that the same condition-aware acquisition principle improves annotation efficiency and robustness across imaging conditions. These results establish a modality-aware active learning strategy for data-efficient SPL perception, with the potential to extend to a broader range of downstream tasks. The dataset and source code are publicly available at
\url{https://github.com/lxy-bee/SinglePhoton-ActiveLearning/tree/main}.

\end{abstract}


\section{Introduction}

Single-photon LiDAR (SPL) is an emerging sensing modality that detects and time-stamps individual photons using highly sensitive single-photon avalanche diodes (SPADs)~\cite{kirmani2014first}. Unlike conventional active sensing systems that rely on dense signal returns, SPL can operate in photon-starved regimes and recover useful information from weak reflections. This sensitivity enables long-range sensing, low-reflectivity target imaging, and perception in attenuating environments~\cite{li2021single}. Moreover, SPL measurements record temporal photon-arrival distributions rather than only a single depth return, providing transient information that can support multipath analysis, scattering-media imaging, and non-line-of-sight reasoning~\cite{liu2022photon}.

These properties make SPL attractive for robotic perception in challenging environments, including autonomous driving, aerial robotics, underwater exploration, and search-and-rescue scenarios. For example, long-range and low-return sensing can support early obstacle perception, while sensitivity under fog, dust, smoke, or underwater attenuation can complement conventional sensors~\cite{goyal2025robust,young2025enhancing,muller2023robust}. However, the use of SPL in robotic systems remains limited because reliable semantic perception under realistic sensing variations is still difficult. Unlike standard RGB-D or LiDAR modalities, SPL lacks large-scale annotated datasets~\cite{zhai2022scaling}. In addition, SPL measurements are strongly affected by photon count and signal-to-background ratio~\cite{suonsivu2024time}, so the same scene may appear substantially different under different operating conditions. This makes annotation costly and causes models trained with limited supervision to generalize poorly across sensing regimes.

To address this problem, we propose a condition-aware active learning framework for data-efficient SPL semantic perception. We focus on image-level classification and semantic segmentation under limited labeled data, which represent two foundational semantic perception tasks. Our method incorporates physics-grounded synthetic SPL variants into the active learning loop, allowing each sample to be evaluated under multiple simulated photon conditions. We further introduce a diversity-guided uncertainty-inconsistency sampling strategy that selects samples that are uncertain, representative, and sensitive to condition changes. In this way, annotation effort is directed toward samples that improve robustness across sensing conditions rather than only those that are ambiguous under a single observation. We evaluate the proposed framework on synthetic benchmarks and real-world SPL measurements for classification and semantic segmentation. Experimental results show that our method consistently outperforms representative active learning baselines while requiring fewer labeled samples. We also release a real SPL depth image classification dataset collected under diverse operating conditions. Overall, this work provides a data-efficient semantic front-end for SPL perception, while full robotic-system deployment, including SPL-based mapping, active perception, and closed-loop navigation, remains an important direction for future work.

\section{Related Work}

Active learning (AL) aims to reduce labeling cost by querying only the most informative samples, and is widely used when annotation is expensive or requires expert knowledge~\cite{ren2021survey}. Existing AL methods can be broadly grouped into uncertainty-based and diversity-based approaches. Uncertainty-based methods prioritize samples that the current model finds difficult, using criteria such as entropy~\cite{lewis1994heterogeneous}, margin~\cite{scheffer2001active}, or Bayesian active learning by disagreement (BALD)~\cite{gal2017deep}. Diversity-based methods select representative samples to improve coverage of the data distribution~\cite{sener2018active}, while hybrid methods combine uncertainty and diversity, for example by sampling in gradient-embedding space~\cite{AshZK0A20}. Although these methods provide effective general-purpose acquisition criteria, they usually do not explicitly model variations caused by the sensing process itself.

Some uncertainty-based AL methods estimate informativeness through prediction instability under input perturbations~\cite{kim2021lada}. For example, ViewAL measures viewpoint-induced prediction entropy for multi-view semantic segmentation~\cite{siddiqui2020viewal}, while consistency-based semi-supervised AL uses disagreement across augmented inputs to guide annotation~\cite{gao2020consistency}. AL has also been explored in robotics and LiDAR perception, including methods based on spatial structure, scene-level diversity, and gradient-based informativeness~\cite{meyer2019importance,schmidt2024generalized,unal2023discwise,sheikh2020gradient}. These studies demonstrate the usefulness of AL for perception tasks, but most of them assume relatively fixed sensing conditions and do not directly account for photon-level imaging variations.

This limitation is important for single-photon LiDAR (SPL), where perception performance can change significantly with photon budget, background illumination, and target reflectance. In this work, we study condition-aware active learning for SPL classification and semantic segmentation. Instead of relying only on generic uncertainty, diversity, or augmentation-induced disagreement, we synthesize SPL variants under different imaging conditions and use the resulting prediction variation as a physics-grounded acquisition signal. The proposed strategy therefore selects samples that are uncertain or representative, while also being sensitive to realistic SPL sensing variations.

\section{Preliminary: Single-Photon Imaging System and Method}
\label{section2}

\begin{figure}[!t]
    \centering \includegraphics[width=0.95\columnwidth,height=0.2\textheight,keepaspectratio]{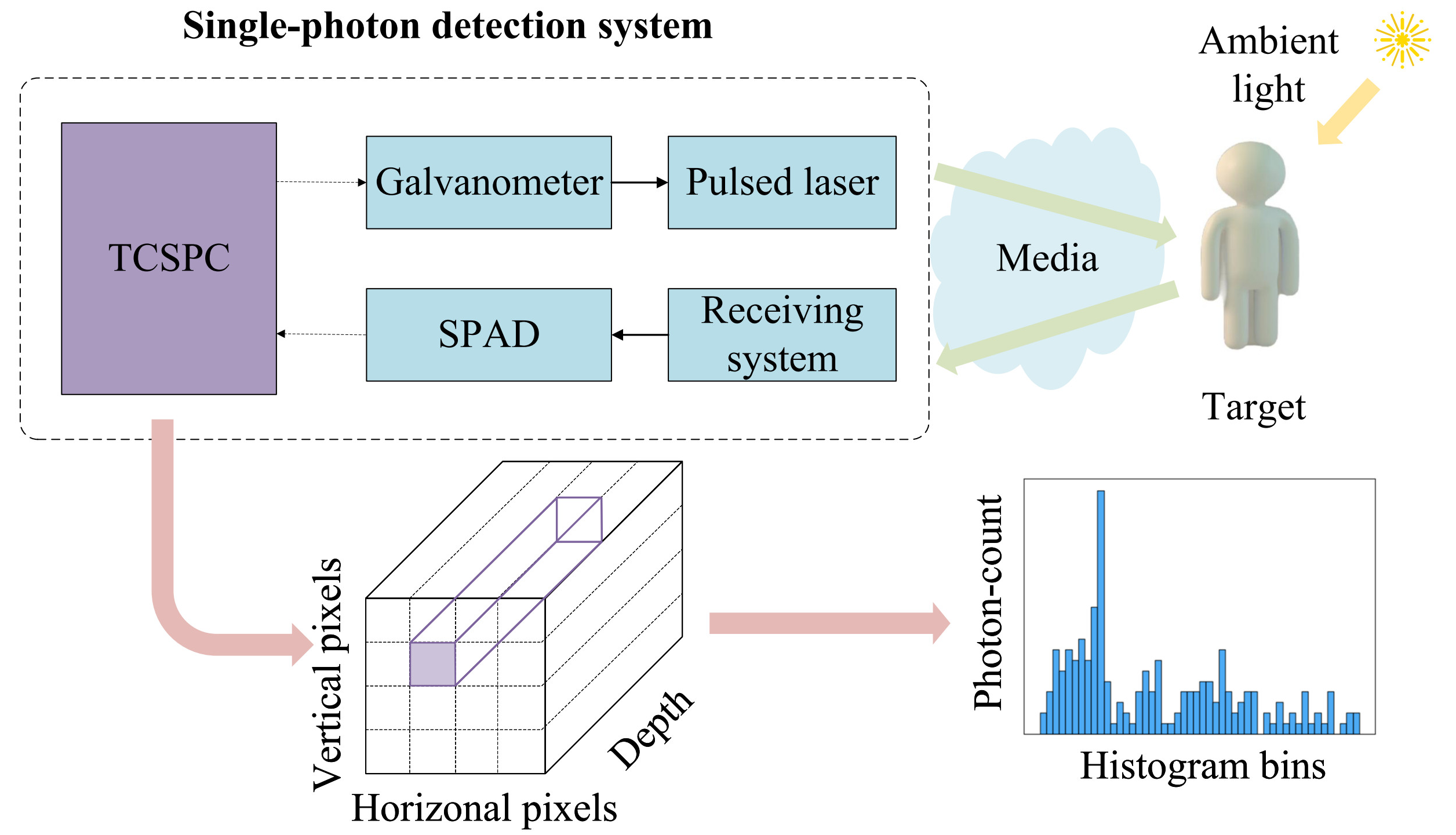}
    \caption{Single-Photon Imaging System Components.} 
    \label{fig:system}     
\end{figure}

The SPL integrates a single-photon detector, pulsed laser, micro-electro-mechanical systems (MEMS) scanner, and time-correlated single-photon counting (TCSPC) module to achieve high-resolution 3D imaging. Pulsed laser emission is angularly directed via MEMS mirrors for systematic target scanning. Reflected photon signals are captured by the single-photon detector, generating ultrafast electrical pulses recorded by the TCSPC module with picosecond precision, as shown in Figure \ref{fig:system}.

The photon-counting histogram acquired by the LiDAR system is modeled as a three-dimensional tensor
$\mathbf{N} \in \mathbb{Z}_+^{N_r \times N_c \times T}$, where $N_r \times N_c$ represents the spatial resolution and $T$ denotes the number of temporal bins. 
Each entry $n_{p,t}$ represents the number of photons detected at pixel $p$ during the $t$-th time bin.

We model these photon counts as independent Poisson random variables:
\[
n_{p,t} \sim \operatorname{Poisson}(\lambda_{p,t}),
\]
where $\lambda_{p,t}$ is the expected photon count for each pixel and time bin.

The expected photon count at each pixel $p$ and time bin $t$ is given by the following expression:
\[
\lambda_{p,t} = g_{p} \left[ a_{p} h(t - t^{d}_{p}) + b_{p} \right],
\]
where $h(\cdot)$ denotes the instrumental response function (IRF), which characterizes how the system reacts to incoming photons over time; $t^{d}_{p}$ represents the time-of-flight bin corresponding to the depth value at pixel $p$; $a_{p}$ is the reflectivity at pixel $p$; $b_{p}$ represents the background photon level at each pixel; $g_{p}$ is a pixel-dependent gain factor.

Assuming statistical independence across pixels and time bins, the negative log-likelihood of the observed photon-counting histogram is:
\[
\mathcal{L}(\mathbf{N}; \mathbf{I}, \mathbf{a}, \mathbf{b})
 = - \sum_{p} \sum_{t=1}^{T} \log \Pr\left(n_{p,t} \mid \lambda_{p,t}\right).
\]

We formulate a maximum likelihood estimation problem by minimizing the negative
log-likelihood of the observed photon-counting histogram $\mathbf{N}$ with respect to depth
$\mathbf{I}$, reflectivity $\mathbf{a}$, and background $\mathbf{b}$, enabling
joint reconstruction of the 3D scene.

\section{Method}
\label{section3}

\subsection{Framework}
\label{3.2}

We propose a condition-aware active learning framework for label-efficient single-photon LiDAR (SPL) semantic perception, covering both image-level classification and semantic segmentation.
The labeled dataset is denoted as
\[
D_{\ell}=\{(I_{\mathrm{obs},i},y_i)\}_{i=1}^{N_{\ell}},
\]
where \(I_{\mathrm{obs},i}\) represents the \(i\)-th observed single-photon depth image
and \(y_i\) denotes its task-specific annotation, i.e., an image-level class label for classification or a pixel-wise semantic mask for segmentation. The unlabeled dataset is denoted as \(D_{\mathrm{u}}=\{I_{\mathrm{obs},i}\}_{i=1}^{N_{\mathrm{u}}}\), which consists of \(N_{\mathrm{u}}\) unlabeled observed images.

To model sensing-condition variations, we generate multiple synthetic SPL variants for each observed image. Specifically, for each observed image \(I_{\mathrm{obs},i}\), we define a sample group as
\[
S_i=\{I_{\mathrm{obs},i}\}\cup\{I_{\mathrm{syn},i,j}\}_{j=1}^{M},
\]
where \(I_{\mathrm{syn},i,j}\) denotes the \(j\)-th synthetic variant associated with \(I_{\mathrm{obs},i}\), and \(M\) denotes the number of synthetic variants. Using this notation, the extended labeled dataset and extended unlabeled dataset are expressed as
\[
D_{\ell}'=\{(S_i,y_i)\}_{i=1}^{N_{\ell}},
\qquad
D_{\mathrm{u}}'=\{S_i\}_{i=1}^{N_{\mathrm{u}}},
\]
respectively.

\begin{figure*}[!t]
    \centering
    \IfFileExists{figs/framworkv3.png}{%
        \includegraphics[scale=0.06]{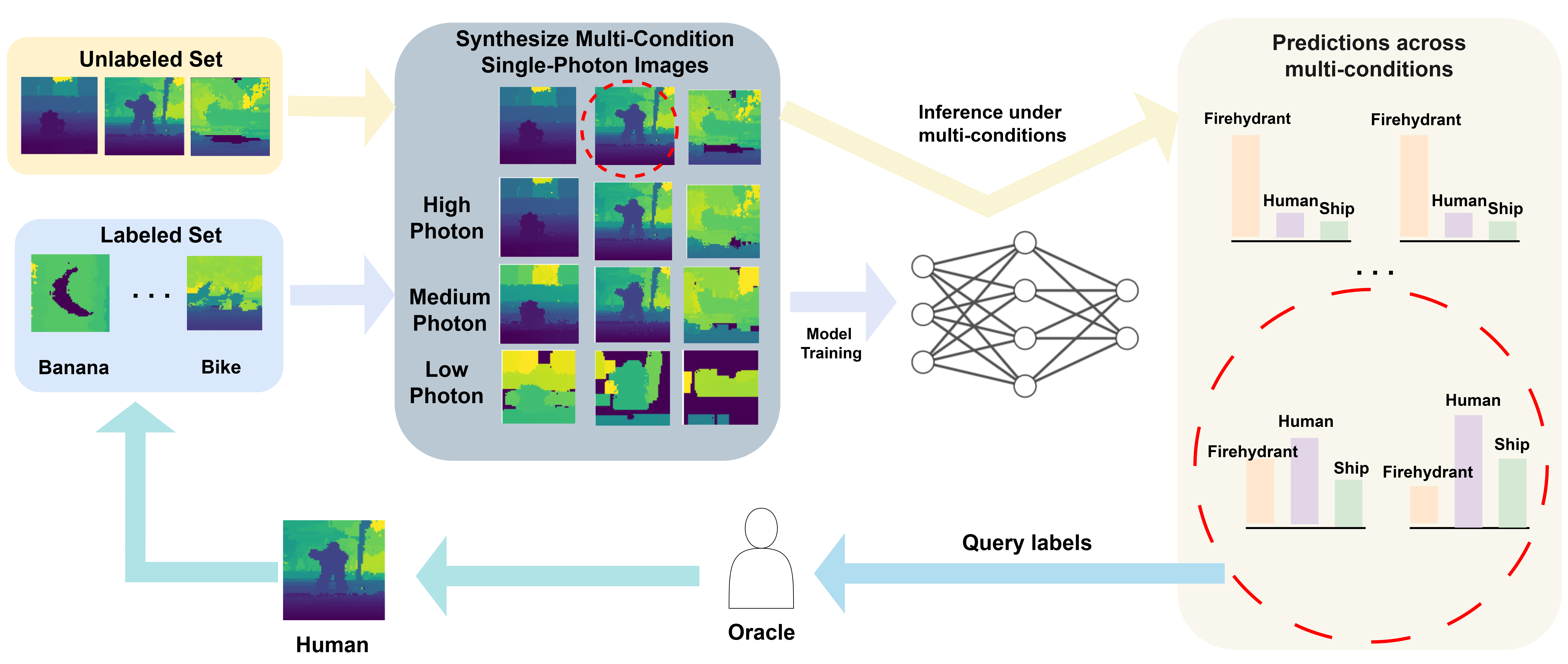}%
    }{%
        \fbox{\parbox[c][35mm][c]{0.85\textwidth}{\centering Placeholder for \texttt{figs/framworkv3.png}}}%
    }
    \caption{Framework of the proposed condition-aware active learning method for SPL perception. In each iteration, the model is trained with labeled data and synthetic SPL variants, selects informative unlabeled sample groups based on condition-aware predictions, and queries the oracle for class labels or pixel-wise masks. The annotated samples are then added to the training set for the next iteration.}
    \label{fig:framework}
\end{figure*}

As shown in Fig.~\ref{fig:framework}, the model \(f_{\theta}(\cdot)\) is first trained on \(D_{\ell}'\). Depending on the task, \(f_{\theta}(\cdot)\) outputs either an image-level class probability distribution or a pixel-wise semantic probability map. During active learning, the model evaluates each unlabeled group \(S_i \in D_{\mathrm{u}}'\). Predictions from the observed image and its synthetic variants are aggregated to measure sample informativeness under different SPL imaging conditions. For classification, uncertainty and inconsistency are computed from image-level class distributions. For segmentation, they are computed from pixel-wise class distributions and then spatially aggregated over valid pixels.

Based on uncertainty, diversity, and condition-induced inconsistency, a batch of informative groups is selected for annotation. The oracle annotates the observed samples, and the same labels or masks are assigned to their corresponding synthetic variants. The newly labeled groups are added to \(D_{\ell}'\), and the model is updated for the next active learning round. This process repeats until the labeling budget is exhausted. In this framework, synthetic SPL variants serve both as augmented training data and as condition-probing views for active sample selection, enabling data-efficient learning under realistic photon-level sensing variations.

\subsection{Generating Multi-Condition Single-Photon Images} 
The uncertainty in single-photon imaging primarily stems from the effects of stochastic photon statistics.  Photon counts follow a Poisson distribution, for which the variance equals the expected photon count. Under low-photon-count conditions, the limited number of detected photons increases the uncertainty of time-of-flight estimation, leading to larger depth measurement errors. Additionally, pixels receiving no detected photons contain insufficient depth information, causing voids in the reconstructed depth map. For the $i$-th observed SPL depth image $I_{\mathrm{obs},i}$, we generate $M$ synthetic variants $\{I_{\mathrm{syn},i,j}\}_{j=1}^{M}$ under different imaging conditions. We synthesize single-photon data through the following two components: (1) signal photon counts $n_{i,j,p}^{\mathrm{sig}}$ with their corresponding TOF bins $\{t_{i,j,p,k}^{\mathrm{sig}}\}_{k=1}^{n_{i,j,p}^{\mathrm{sig}}}$, and (2) background photon counts $n_{i,j,p}^{\mathrm{bkg}}$ with associated TOF bins $\{t_{i,j,p,k}^{\mathrm{bkg}}\}_{k=1}^{n_{i,j,p}^{\mathrm{bkg}}}$. Here, $p\in\mathcal{P}_i$ denotes a valid pixel location in $I_{\mathrm{obs},i}$, $\mathcal{P}_i$ is the valid pixel set, and $j$ denotes the synthetic imaging condition. We first configure the detector's physical parameters, including the time-bin duration $\Delta t$ and the temporal standard deviation $T_{\mathrm{pulse}}$ of the instrumental response. We denote the maximum valid time bin by $T_{\max}$. For each valid pixel $p$, the observed depth value $I_{\mathrm{obs},i}(p)$ is converted into the corresponding time bin: 
\begin{equation} 
 t^{\mathrm{obs}}_{i,p} = \operatorname{round} \left( \frac{2I_{\mathrm{obs},i}(p)} {c\Delta t} \right),  \label{eq:tobs} 
\end{equation} 
where $c$ represents the speed of light.  For each detected signal photon, the arrival bin under synthetic condition $j$ is modeled using a Gaussian pulse profile with mean $t_{i,p}^{\mathrm{obs}}$ and standard deviation $\sigma=T_{\mathrm{pulse}}/\Delta t$ in temporal-bin units: 
\begin{equation} 
 t^{\mathrm{sig}}_{i,j,p,k} = \operatorname{clip}_{[0,T_{\max}]} \left\{ \operatorname{round} \left[ t^{\mathrm{obs}}_{i,p} + \sigma\mathcal{N}(0,1) \right] \right\}, \label{eq:tsig} 
\end{equation} 
where $k=1,\ldots,n_{i,j,p}^{\mathrm{sig}}$, $\operatorname{clip}_{[0,T_{\max}]}(\cdot)$ restricts the sampled arrival bin to the valid temporal range. 

\begin{algorithm}[!htbp] \caption{Generating a Synthetic SPL Variant} \label{alg:generate_simdata} \KwIn{ Observed SPL depth image $I_{\mathrm{obs},i}$, normalized reflectance $\hat A_{\mathrm{obs},i}$, MSPPP $\mu_{i,j}^{\mathrm{sig}}$, time-bin duration $\Delta t$, pulse standard deviation $T_{\mathrm{pulse}}$, maximum valid time bin $T_{\max}$, signal-to-background ratio SBR} \KwOut{Synthetic SPL image $I_{\mathrm{syn},i,j}$} \ForEach{valid pixel $p\in\mathcal{P}_i$}{ \textbf{Compute Time Bin:} Compute $t^{\mathrm{obs}}_{i,p}$ by Eq.~\eqref{eq:tobs}.\; \textbf{Generate Signal Photons:} Generate $n_{i,j,p}^{\mathrm{sig}}$ by Eq.~\eqref{eq:nsig} and $\{t_{i,j,p,k}^{\mathrm{sig}}\}_{k=1}^{n_{i,j,p}^{\mathrm{sig}}}$ by Eq.~\eqref{eq:tsig}. }\; \textbf{Generate Background Noise:}  Generate $n_{i,j,p}^{\mathrm{bkg}}$ by Eq.~\eqref{eq:nback} and $\{t_{i,j,p,k}^{\mathrm{bkg}}\}_{k=1}^{n_{i,j,p}^{\mathrm{bkg}}}$ by Eq.~\eqref{eq:tback}.\;  \textbf{Generate Simulated Single-Photon Raw Data:} Combine the generated signal and background detection events to obtain the raw single-photon measurement $\mathcal{R}_{i,j}$.\; \textbf{Generate Synthetic Single-Photon Image:}  Input the raw single-photon data into the sparse single-photon imaging network, $I_{\mathrm{syn},i,j}\leftarrow \mathrm{SSPI}(\mathcal{R}_{i,j})$. \; \Return{$I_{\mathrm{syn},i,j}$}. \end{algorithm}

The number of signal photons \(n_{i,j,p}^{\mathrm{sig}}\) follows a Poisson random process described by: 
\begin{equation}  n_{i,j,p}^{\mathrm{sig}} \sim \operatorname{Poisson} \left( \mu_{i,j}^{\mathrm{sig}} \hat A_{\mathrm{obs},i}(p) \right), \label{eq:nsig} 
\end{equation}
where \(\mu_{i,j}^{\mathrm{sig}}\) represents the mean signal photons per pixel (MSPPP) for the \(j\)-th synthetic condition and  $\hat A_{\mathrm{obs},i}(p)$ denotes the reflectance at pixel $p$, normalized to have unit mean over the valid pixel set $\mathcal{P}_i$.  The background photon count \(n_{i,j,p}^{\mathrm{bkg}}\) is also generated using a Poisson distribution based on the predefined signal-to-background ratio (SBR): \begin{equation} n_{i,j,p}^{\mathrm{bkg}} \sim \operatorname{Poisson} \left( \frac{ \operatorname{mean}_{p\in\mathcal{P}_i} \left[ \mu_{i,j}^{\mathrm{sig}} \hat A_{\mathrm{obs},i}(p) \right] } {\mathrm{SBR}} \right).  \label{eq:nback} \end{equation} 
Here, the expected signal-photon count, rather than a realized random signal count, is used to determine the background Poisson rate. According to~\cite{rapp2017few}, the background detection time follows a uniform distribution: 
\begin{equation}  t_{i,j,p,k}^{\mathrm{bkg}} \sim \operatorname{Unif} \left( \{0,1,\ldots,T_{\max}\} \right), \quad k=1,\ldots,n_{i,j,p}^{\mathrm{bkg}}, \label{eq:tback}
\end{equation} 
where \(T_{\max}\) represents the maximum valid time bin. By mixing the generated signal photons with background photons according to Eqs.~\eqref{eq:nsig}--\eqref{eq:tback}, the raw single-photon measurement $\mathcal{R}_{i,j}$ can be obtained. The detailed implementation steps are outlined in Algorithm~\ref{alg:generate_simdata}. 

Algorithm~\ref{alg:generate_simdata} defines the synthetic imaging variants (SIV) module. By systematically varying imaging conditions such as MSPPP and SBR, SIV generates multiple realistic variants $\{I_{\mathrm{syn},i,j}\}_{j=1}^{M}$ from a single observed SPL image $I_{\mathrm{obs},i}$. These variants capture the intrinsic stochasticity of photon measurements and provide the necessary input for estimating both uncertainty and prediction inconsistency in the subsequent active learning process.

\begin{algorithm}[!htbp]
\vspace{5pt}
\caption{Active Learning with DUIS}
\label{alg:active_learning}
\KwIn{
Extended labeled set $D_{\ell}'$; extended unlabeled set $D_{\mathrm{u}}'$; model $f_{\theta}$; maximum iterations $T$; batch size $N_{\mathrm{batch}}$; candidate-pool size $N_{\mathrm{cand}}$; trade-off weight $\alpha$
}
\KwOut{Updated extended labeled set $D_{\ell}'$ and updated model $f_{\theta}$}

\For{$t=1$ \KwTo $T$}{
Train $f_{\theta}$ on $D_{\ell}'$.\;

Extract group features $\phi_i$ for all sample groups $S_i\in D_{\mathrm{u}}'$ and apply $k$-means clustering to form the candidate pool $\mathcal{C}$ with $N_{\mathrm{cand}}$ groups.\;

\ForEach{$S_i\in\mathcal{C}$}{
Compute the task-specific mean prediction $\bar q_i(u)$ by Eq.~\eqref{eq:task_prediction_mean}, where $u\in\Omega_i$ denotes the task unit.\;

Compute the margin score $\gamma_i$ by Eq.~\eqref{eq:margin_mean}.\;

Compute the condition-inconsistency score $d_i$ by Eq.~\eqref{eq:condition_inconsistency}.\;
}

Normalize $\{\gamma_i\}_{S_i\in\mathcal{C}}$ and $\{d_i\}_{S_i\in\mathcal{C}}$ by Eq.~\eqref{eq:normalization} to obtain $\widetilde \gamma_i$ and $\widetilde d_i$.\;

Compute $\operatorname{UIS}_{\alpha}(S_i)$ for each $S_i\in\mathcal{C}$ by Eq.~\eqref{eq:uis_score}.\;

Select the top $N_{\mathrm{batch}}$ groups as the query batch $\mathcal{B}$, obtain their annotations from the oracle, and update $D_{\ell}'$ and $D_{\mathrm{u}}'$.\;
}
\Return{$D_{\ell}'$, $f_{\theta}$}
\end{algorithm}

\subsection{Active Sample Selection}
\label{sec:active_sample_selection}

After training \(f_{\theta}(\cdot)\) on the current extended labeled set \(D_{\ell}'\), we select a batch of informative and diverse sample groups from the extended unlabeled set \(D_{\mathrm{u}}'\) for annotation. Here, \(D_{\ell}'\) contains labeled sample groups and \(D_{\mathrm{u}}'\) contains unlabeled sample groups, following the notation in Sec.~\ref{3.2}. For each unlabeled group \(S_i\in D_{\mathrm{u}}'\), we first extract penultimate-layer features and compute a group feature
\[
\phi_i=(M+1)^{-1}\sum_{m=0}^{M}\phi_{\theta}(I_i^{(m)}),
\]
where \(\phi_{\theta}(\cdot)\) denotes the feature extractor induced by \(f_{\theta}(\cdot)\), \(S_i=\{I_i^{(m)}\}_{m=0}^{M}\) is the group associated with the \(i\)-th observed sample, \(I_i^{(0)}=I_{\mathrm{obs},i}\) is the original observed SPL input, and \(I_i^{(m)}=I_{\mathrm{syn},i,m}\) for \(m=1,\ldots,M\) are its \(M\) synthetic condition variants. We apply \(k\)-means clustering to the group features \(\{\phi_i\}_{S_i\in D_{\mathrm{u}}'}\) and select the groups closest to the cluster centers to form a diverse candidate pool \(\mathcal{C}\subset D_{\mathrm{u}}'\), where \(|\mathcal{C}|=N_{\mathrm{cand}}\) and \(N_{\mathrm{cand}}\) is the candidate-pool size.

To use the same acquisition rule for both image-level classification and semantic segmentation, we define a task unit $u\in\Omega_i$, where $\Omega_i$ is the set of valid prediction units for sample group $S_i$. For classification, $\Omega_i$ contains one image-level unit. For semantic segmentation, $\Omega_i$ contains valid pixels or points. Let $F_{\theta}$ denote the task-specific probability output of $f_{\theta}$, and let $C$ be the number of classes. For the \(m\)-th SPL condition, the prediction of unit \(u\) and its group average are defined as
\begin{equation}
\begin{aligned}
q_i^{(m)}(u)
&=
F_{\theta}\!\left(I_i^{(m)}\right)(u)
\in \Delta^{C-1}, \\
\bar q_i(u)
&=
\frac{1}{M+1}
\sum_{m=0}^{M} q_i^{(m)}(u),
\end{aligned}
\label{eq:task_prediction_mean}
\end{equation}
where \(\Delta^{C-1}\) denotes the \((C-1)\)-dimensional probability simplex, \(q_i^{(m)}(u)\) is the class-probability vector under the \(m\)-th condition, and \(\bar q_i(u)\) is the mean prediction over the observed input and all synthetic variants.

We compute margin-based uncertainty from the averaged prediction. Let \(c_i^{(1)}(u)\) and \(c_i^{(2)}(u)\) denote the classes with the largest and second-largest probabilities in \(\bar q_i(u)\), respectively. The group margin is denoted by \(\gamma_i\):
\begin{equation}
\gamma_i=
\frac{1}{|\Omega_i|}
\sum_{u\in\Omega_i}
\left[
\bar q_i\!\left(u,c_i^{(1)}(u)\right)
-
\bar q_i\!\left(u,c_i^{(2)}(u)\right)
\right],
\label{eq:margin_mean}
\end{equation}
where \(|\Omega_i|\) is the number of valid task units in \(S_i\). A smaller margin \(\gamma_i\) indicates higher predictive uncertainty.

We measure condition-induced inconsistency by the average KL divergence between each condition-specific prediction and the mean prediction:
\begin{equation}
d_i=
\frac{1}{|\Omega_i|}
\sum_{u\in\Omega_i}
\frac{1}{M+1}
\sum_{m=0}^{M}
D_{\mathrm{KL}}\!\left(
q_i^{(m)}(u)
\,\middle\|\,
\bar q_i(u)
\right),
\label{eq:condition_inconsistency}
\end{equation}
where \(D_{\mathrm{KL}}(\cdot\|\cdot)\) is the Kullback--Leibler divergence and \(d_i\) denotes the condition-inconsistency score of \(S_i\). For classification, \(|\Omega_i|=1\), so \(d_i\) reduces to image-level group inconsistency. For semantic segmentation, \(d_i\) is averaged over valid pixels or points.

Because \(\gamma_i\) and \(d_i\) have different numerical scales, we normalize them within the current candidate pool \(\mathcal{C}\):
\begin{equation}
\begin{aligned}
\widetilde d_i
&=
\frac{
d_i-\min_{S_j\in\mathcal{C}} d_j
}{
\max_{S_j\in\mathcal{C}} d_j
-
\min_{S_j\in\mathcal{C}} d_j
+\epsilon
}, \\
\widetilde \gamma_i
&=
\frac{
\gamma_i-\min_{S_j\in\mathcal{C}} \gamma_j
}{
\max_{S_j\in\mathcal{C}} \gamma_j
-
\min_{S_j\in\mathcal{C}} \gamma_j
+\epsilon
}.
\end{aligned}
\label{eq:normalization}
\end{equation}
where \(\widetilde d_i\) and \(\widetilde \gamma_i\) are the normalized inconsistency and margin scores, and \(\epsilon\) is a small constant for numerical stability. Since a smaller margin corresponds to higher uncertainty, we use \(1-\widetilde \gamma_i\) as the normalized uncertainty score.

The final uncertainty-inconsistency score is defined as
\begin{equation}
\operatorname{UIS}_{\alpha}(S_i)
=
\alpha\widetilde d_i
+
(1-\alpha)(1-\widetilde \gamma_i),
\qquad
\alpha\in[0,1],
\label{eq:uis_score}
\end{equation}
where \(\alpha\) controls the trade-off between condition inconsistency and predictive uncertainty. We set \(\alpha=0.5\) unless otherwise specified, and query the top \(N_{\mathrm{batch}}\) groups in \(\mathcal{C}\) according to \(\operatorname{UIS}_{\alpha}\), where \(N_{\mathrm{batch}}\) is the annotation budget of the current active learning round.

The selected batch \(\mathcal{B}\) is sent to the oracle for task-specific annotation \(y_i\), where \(y_i\) denotes a class label for image-level classification or a dense semantic mask for semantic segmentation. The pools are updated as
\[
D_{\ell}\leftarrow D_{\ell}\cup\{(I_{\mathrm{obs},i},y_i):S_i\in\mathcal{B}\},
\]
\[
D_{\ell}'\leftarrow D_{\ell}'\cup\{(S_i,y_i):S_i\in\mathcal{B}\},
\qquad
D_{\mathrm{u}}'\leftarrow D_{\mathrm{u}}'\setminus\mathcal{B}.
\]
This gives our diversity-guided uncertainty-inconsistency sampling strategy, DUIS, where clustering encourages batch diversity and \(\operatorname{UIS}_{\alpha}\) selects samples that are both uncertain and sensitive to SPL condition changes.

\section{Experiments and Results}
\label{section4}
We conducted systematic experiments on both synthetic datasets and real-world single-photon LiDAR datasets. 

\subsection{Experiments on the Synthetic Single-Photon Classification Dataset}
\label{sec:exp_syn_cls}

\subsubsection{Experimental Setup}

\textbf{Dataset:} The synthetic SPL dataset contains 11 categories generated from depth images acquired from the RGB-D Object Dataset~\cite{lai2011large} and an Intel RealSense D435i camera. The original $640\times360$ depth images are downsampled to $128\times72$ to approximate the spatial resolution of SPL measurements. Each category is randomly divided into 70\% training and 30\% testing samples.

\textbf{Implementation Details:} We use a randomly initialized ResNet-18 trained for 200 epochs with SGD, a momentum of 0.9, an initial learning rate of 0.1, a batch size of 512, and cosine learning-rate annealing. Each observed sample is associated with four synthetic photon-condition variants, i.e., $M=4$. Active learning is performed for seven rounds with 50 samples queried per round, resulting in a total budget of 350 samples. We set $N_{\mathrm{cand}}=500$ and report the mean accuracy and standard deviation over three runs.

\textbf{Baselines:} We compare DUIS with Random sampling, Entropy~\cite{lewis1994heterogeneous}, Margin~\cite{scheffer2001active}, Coreset~\cite{sener2018active}, and BADGE~\cite{AshZK0A20}. These methods respectively represent random selection, predictive uncertainty, representation coverage, and joint uncertainty-diversity sampling.

\subsubsection{Results}
\begin{figure}[!t]
\centering
\includegraphics[scale=0.15]{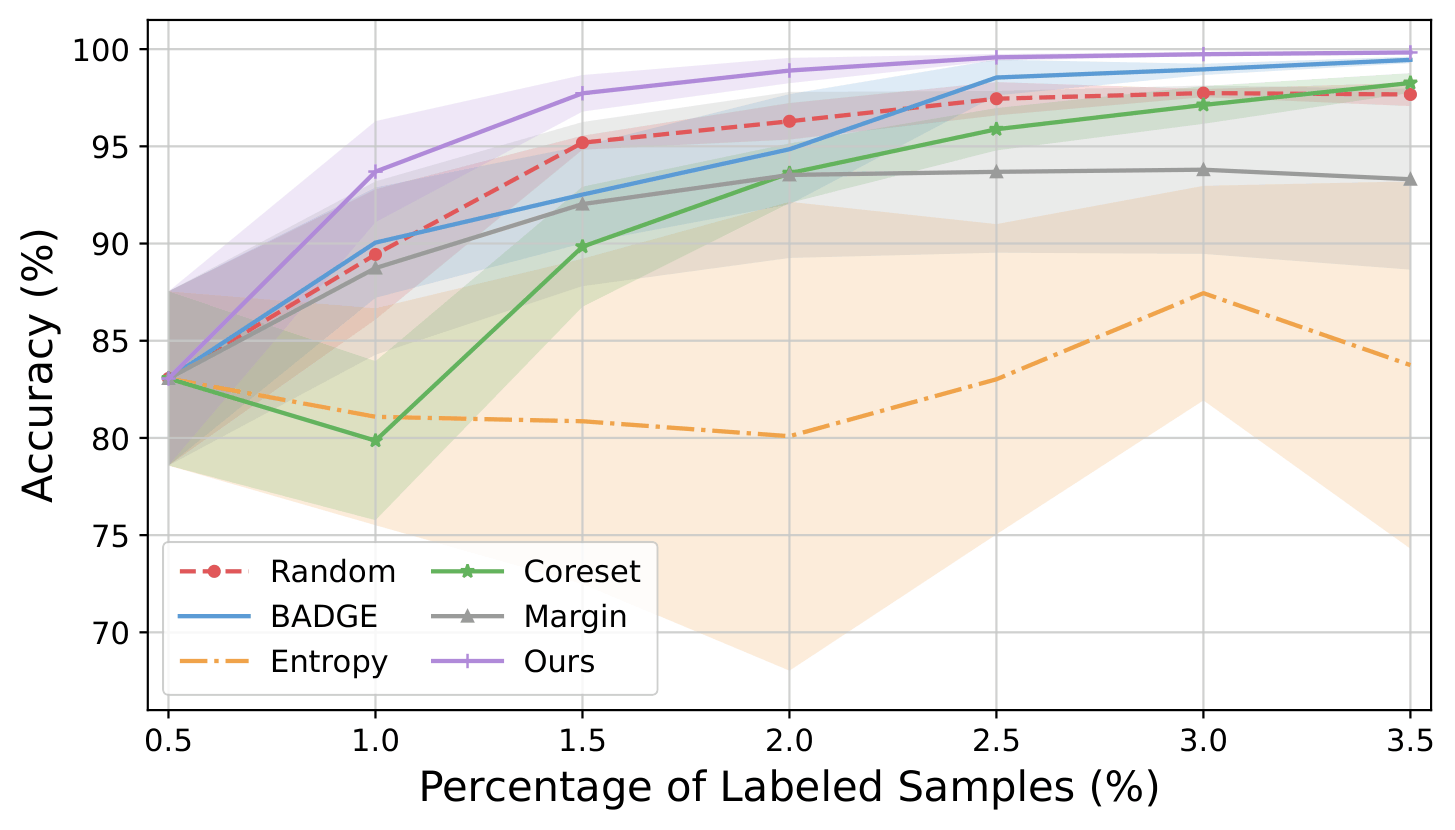}
\caption{Classification accuracy versus labeling ratio on the synthetic single-photon LiDAR dataset.}
\label{fig:11cls}
\end{figure}

Figure~\ref{fig:11cls} shows that classification accuracy steadily improves as the labeling budget increases for all methods. DUIS consistently achieves the highest accuracy across the evaluated budgets, with a more pronounced advantage in the low-label regime. This indicates that DUIS is able to identify highly informative samples at an early stage of active learning, allowing the model to achieve strong performance with substantially fewer annotations. The advantage remains as the labeling budget grows, demonstrating the stability of the proposed acquisition strategy over multiple active learning rounds.  Overall, these results demonstrate improved sample-selection efficiency and better use of the available annotation budget compared with the competing baselines.

\subsection{Experiments on Synthetic Single-Photon Semantic Segmentation Dataset}
\label{sec:exp_seg}

We further evaluate our methods on the  VKITTI 2 dataset~\cite{cabon2020vkitti2} to examine its applicability to dense prediction. The dataset contains 7002 training samples and 3000 test samples. We train PointNet$++$ using point-wise cross entropy and add 100 labeled samples in each active learning round. All results are averaged over three runs with seeds 1, 2, and 3.

\begin{figure}[!t]
\centering
\includegraphics[scale=0.14]{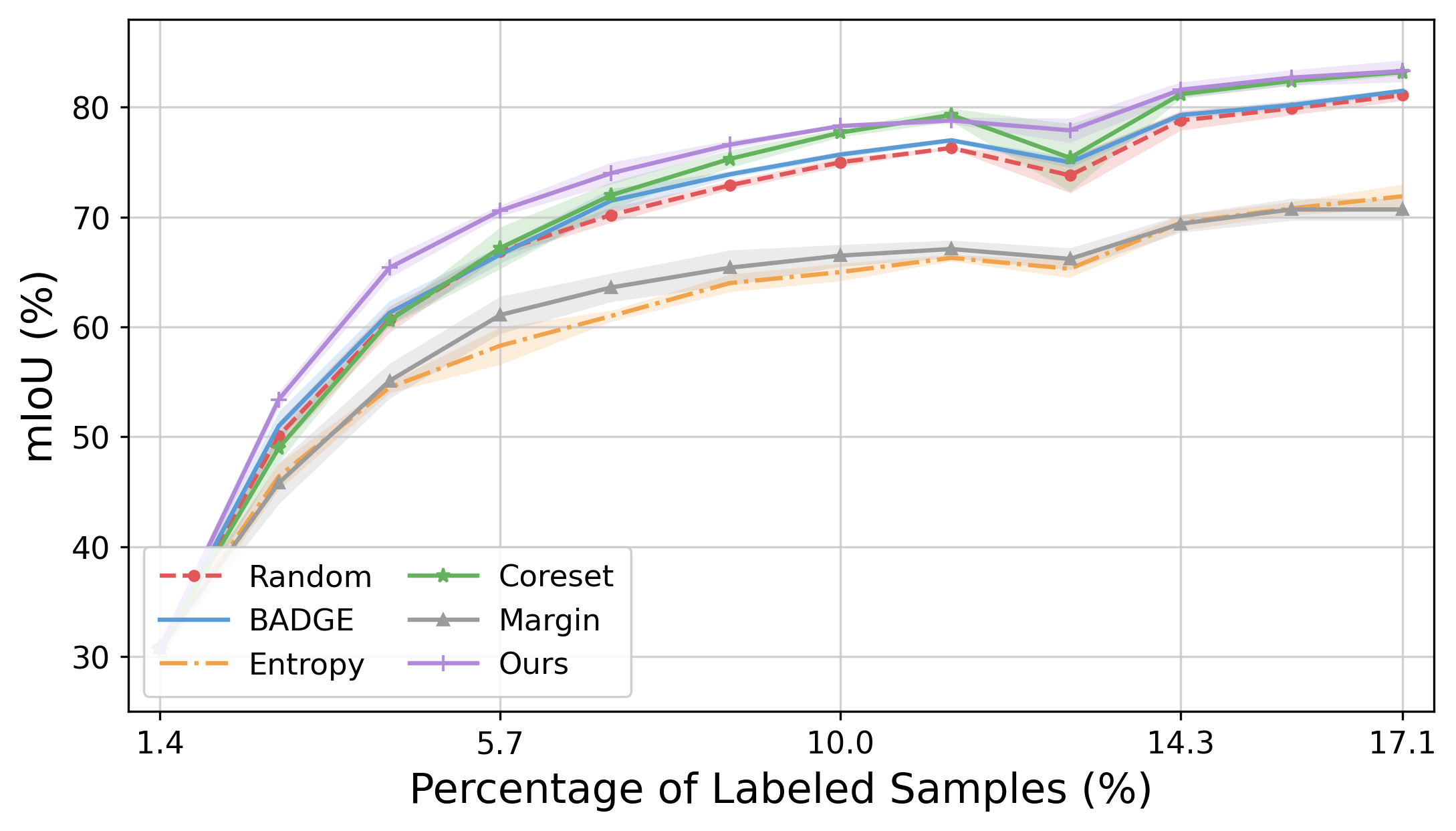}
\caption{Semantic segmentation mIoU versus labeling ratio on the synthetic single-photon LiDAR dataset.}

\label{fig:seg_miou_curve}
\end{figure}

\begin{table}[!t] \centering \caption{Semantic segmentation comparison on the simulated VKITTI point-cloud dataset under different labeling budgets.} \label{tab:segmentation_comparison_r2c1} \setlength{\tabcolsep}{5pt} \renewcommand{\arraystretch}{1.12} \begingroup \begin{tabular}{lccccc} \toprule Method & 2.9\% & 5.7\% & 8.6\% & 12.9\% & 17.1\% \\ \midrule BADGE & 51.0$\pm$1.3 & 66.6$\pm$0.8 & 73.9$\pm$0.3 & 75.0$\pm$0.5 & 81.5$\pm$0.2 \\ Entropy & 46.4$\pm$1.2 & 58.3$\pm$1.7 & 64.0$\pm$0.8 & 65.3$\pm$0.8 & 71.9$\pm$1.1 \\ Margin & 45.8$\pm$1.9 & 61.1$\pm$1.7 & 65.4$\pm$1.6 & 66.2$\pm$1.0 & 70.7$\pm$1.0 \\ Coreset & 49.0$\pm$1.3 & 67.2$\pm$1.9 & 75.3$\pm$0.8 & 75.4$\pm$3.1 & 83.2$\pm$0.3 \\  Random & 50.1$\pm$1.3 & 66.9$\pm$0.4 & 72.9$\pm$0.4 & 73.8$\pm$1.6 & 81.1$\pm$0.5 \\ DUIS (Ours) & \textbf{53.4$\pm$0.7} & \textbf{70.6$\pm$0.5} & \textbf{76.6$\pm$0.4} & \textbf{77.9$\pm$1.1} & \textbf{83.3$\pm$1.0} \\ \bottomrule \end{tabular} \endgroup \end{table}

As shown in Fig.~\ref{fig:seg_miou_curve} and summarized in Table~\ref{tab:segmentation_comparison_r2c1}, DUIS achieves the highest mIoU at all evaluated labeling budgets. The maximum relative improvement over the strongest baseline is approximately 5.1\%, indicating improved label efficiency for point-wise semantic prediction.

\begin{table}[!t]
\centering
\caption{Semantic segmentation performance on the VKITTI dataset at 12.9\% labeled samples under different MSPPP splits.}
\label{tab:sim_seg_msppp_12}

\begingroup
\begin{tabular}{lccc}
\toprule
Method & High MSPPP & Medium MSPPP & Low MSPPP \\
\midrule
Entropy & 65.4 $\pm$ 0.6 & 65.8 $\pm$ 1.1 & 64.8 $\pm$ 0.7 \\
Margin & 66.9 $\pm$ 1.1 & 66.1 $\pm$ 0.7 & 65.7 $\pm$ 1.0 \\
Coreset & 75.2 $\pm$ 3.2 & 75.8 $\pm$ 2.9 & 75.1 $\pm$ 3.1 \\
BADGE & 74.7 $\pm$ 0.4 & 75.4 $\pm$ 0.5 & 74.6 $\pm$ 0.7 \\
Random & 73.5 $\pm$ 1.8 & 74.0 $\pm$ 1.1 & 74.0 $\pm$ 1.6 \\
DUIS (Ours) & \textbf{77.4 $\pm$ 1.2} & \textbf{78.1 $\pm$ 1.2} & \textbf{77.9 $\pm$ 1.0} \\
\bottomrule
\end{tabular}
\endgroup

\end{table}

Table~\ref{tab:sim_seg_msppp_12} shows that DUIS also achieves the highest mIoU under High, Medium, and Low MSPPP conditions, outperforming the strongest baseline by 2.2-2.8 mIoU points. This result indicates consistent performance across the evaluated photon-density regimes.




\subsection{Experiments on the Real Single-Photon Dataset}
\subsubsection{System Description and Dataset}

\begin{figure}[!t]
    \centering
    \includegraphics[scale=0.5]{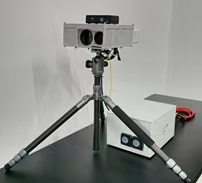}
    \caption{Our single-photon LiDAR system.}
    \label{fig:system2}
\end{figure}

We use a custom single-photon LiDAR system (Fig.~\ref{fig:system2}) with access to raw photon timing histograms for each pixel. The system uses 1~MHz nanosecond laser pulses, an InGaAs/InP SPAD detector, 15~ps timing resolution, and MEMS-based angular scanning to acquire $64\times64$ depth frames. Its timing jitter is about 520~ps, corresponding to an effective depth resolution of approximately 10~cm within 120~m.

\begin{figure}[!t]
    \centering
    \setlength{\tabcolsep}{1pt}
    \renewcommand{\arraystretch}{0}
    \begin{tabular}{ccc}
        \includegraphics[width=0.3\columnwidth]{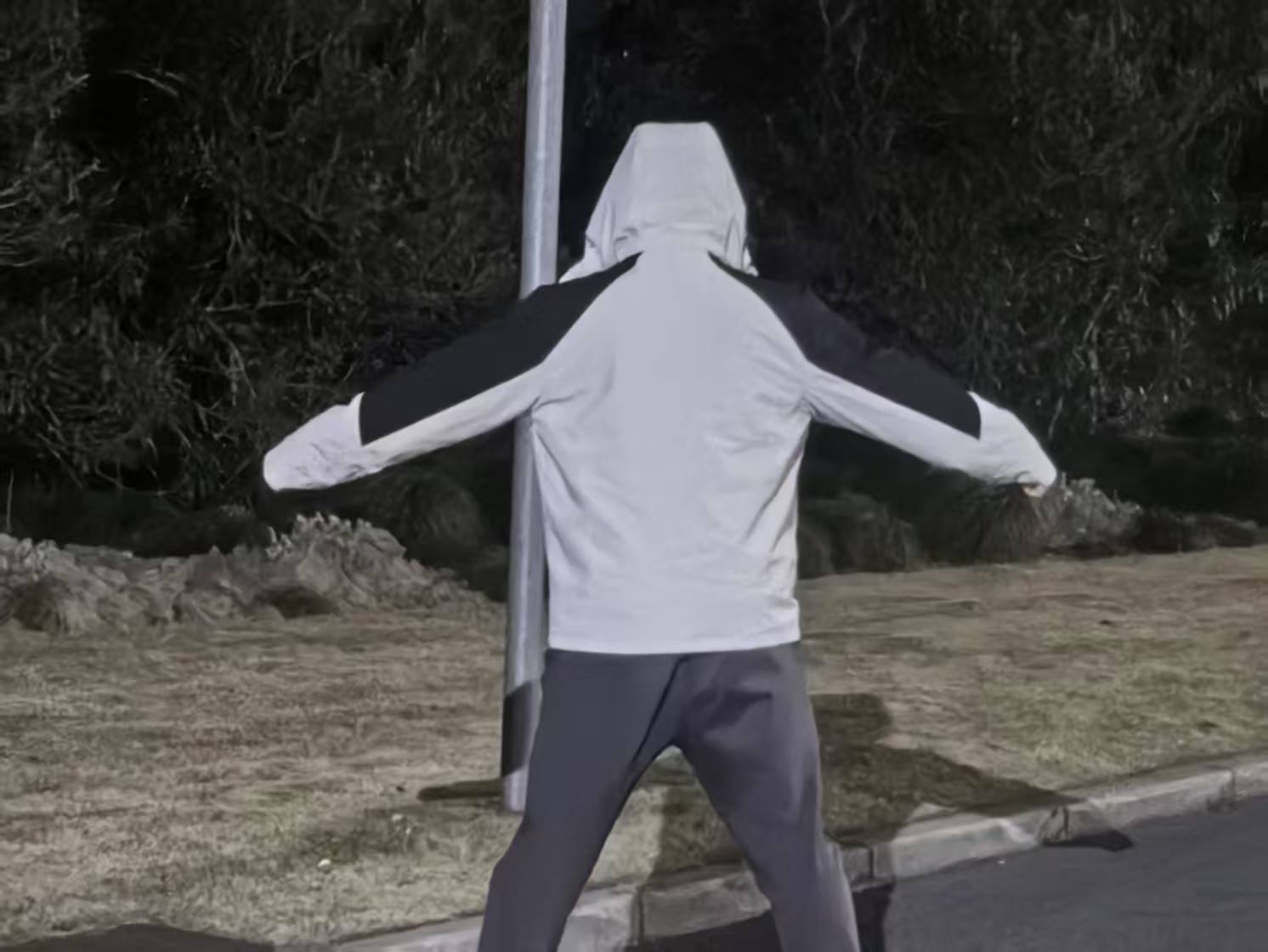} &
        \includegraphics[width=0.3\columnwidth]{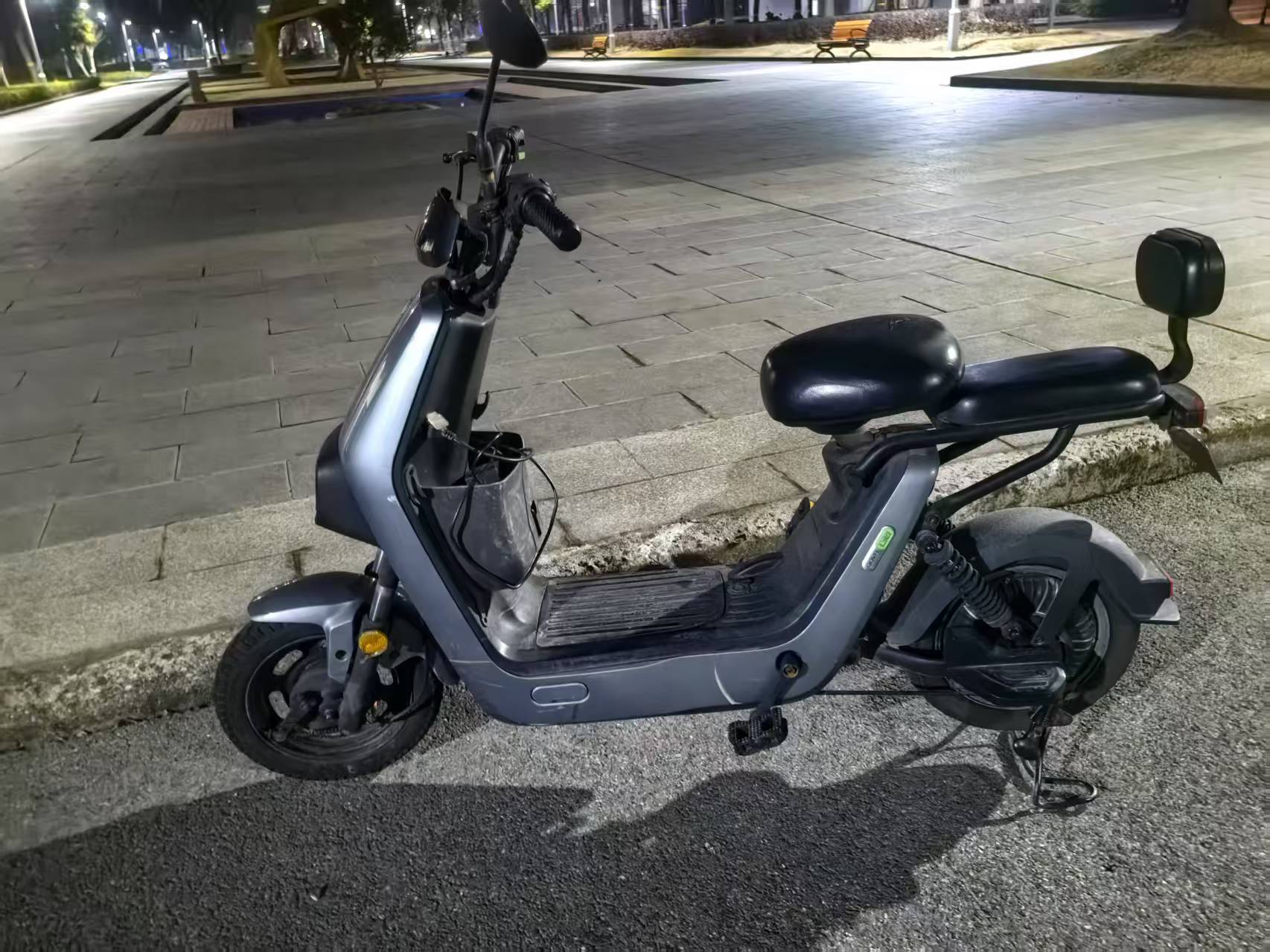}  &
        \includegraphics[width=0.3\columnwidth]{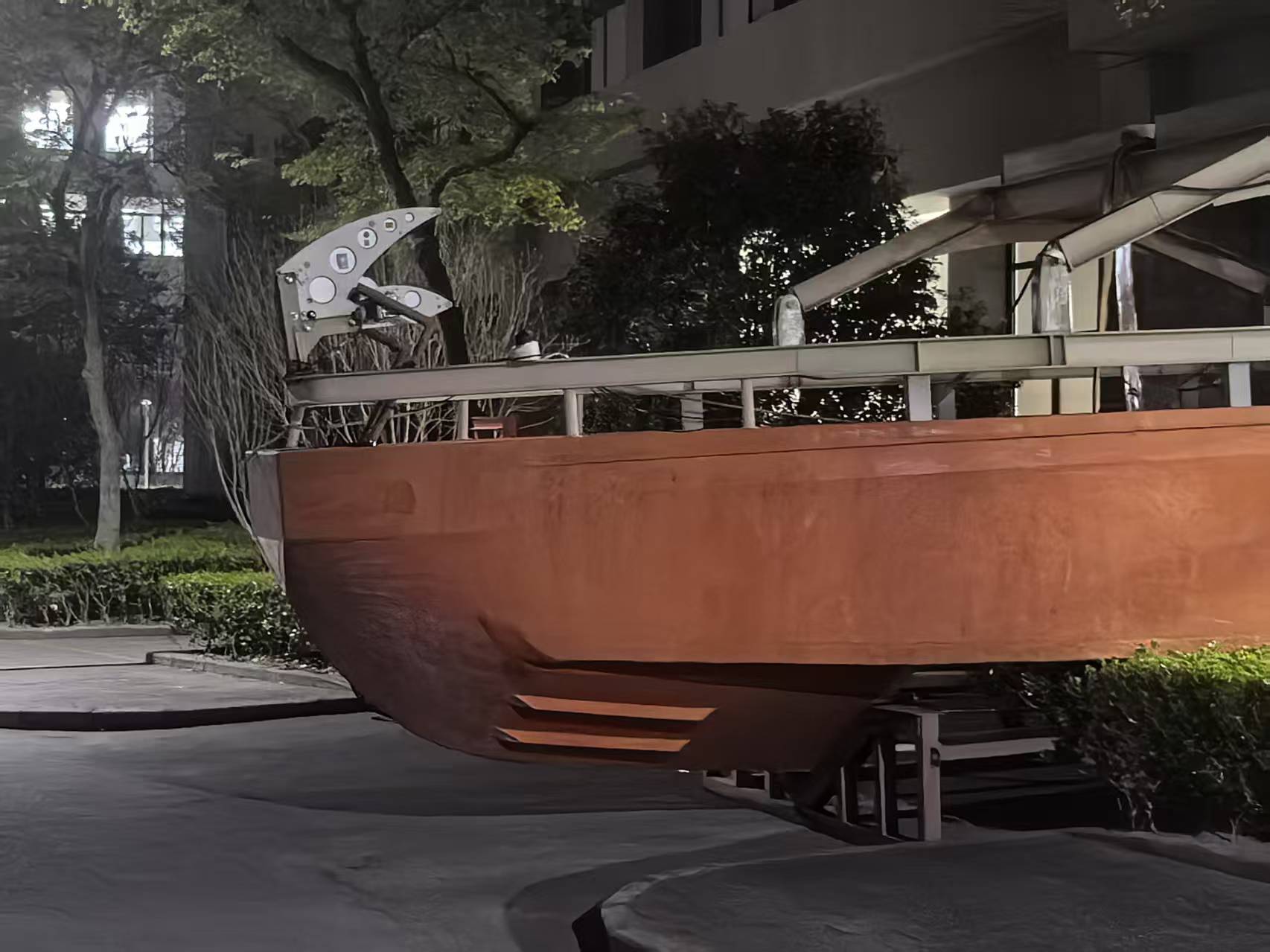}  \\
        \includegraphics[width=0.3\columnwidth]{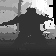}    &
        \includegraphics[width=0.3\columnwidth]{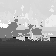}     &
        \includegraphics[width=0.3\columnwidth]{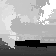}     \\
    \end{tabular}
    \caption{Examples of RGB images and reconstructed single-photon images.}
    \label{fig:realSP}
\end{figure}

\begin{figure}[!t]
    \centering 
    \includegraphics[scale=0.53]{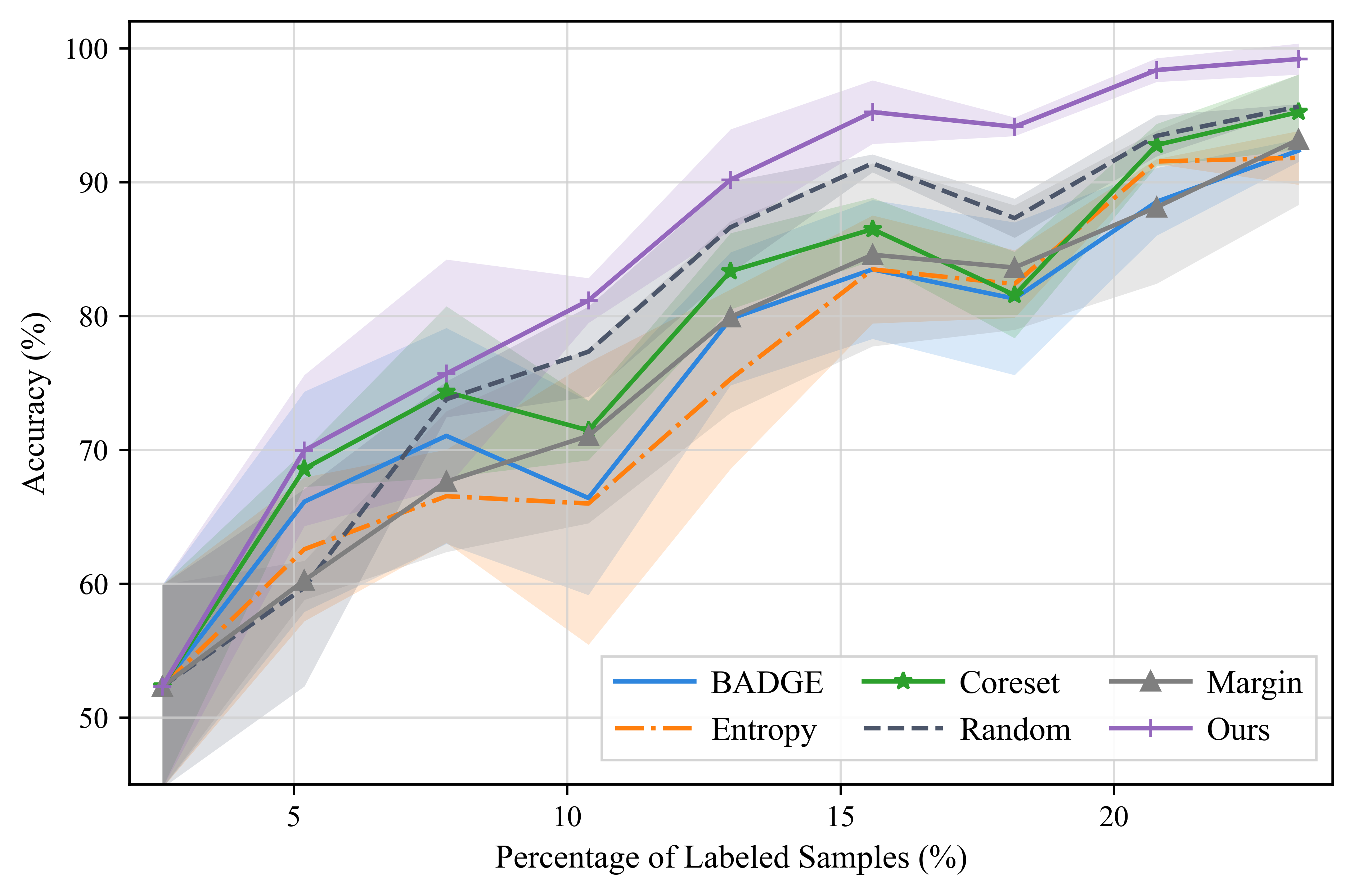}
    \caption{Classification accuracy versus labeling ratio on the real-world single-photon LiDAR dataset.}

    \label{fig:real_acc}                         
\end{figure}

\textbf{Real-world Single-photon Dataset:}
We collected 1,100 single-photon frames from 11 object categories, with 100 frames per category. The data follow the same geometric settings as the synthetic dataset, while the detection distance is adjusted according to target size. Raw photon-counting data are reconstructed by SSPI~\cite{yao2022dynamic} into $64\times64$ single-photon images. Examples are shown in Fig.~\ref{fig:realSP}. For evaluation, each category is randomly split into 70\% training samples and 30\% testing samples.

\textbf{Implementation Details:}
The experimental settings are the same as those used in the synthetic classification experiment, except that active sampling is performed over 10 rounds with 20 samples selected per round.

\subsubsection{Experimental Results and Analysis}

\begin{table}[!t]
\centering
\caption{Classification performance on the real SPL dataset at 7.3\% labeled samples under different MSPPP splits.}
\label{tab:real_cls_msppp}
\begingroup
\begin{tabular}{lccc}
\toprule
Method & High MSPPP & Medium MSPPP & Low MSPPP \\
\midrule
BADGE & $66.4 \pm 8.9$ & $73.1 \pm 9.5$ & $68.4 \pm 6.9$ \\
Entropy & $66.0 \pm 12.9$ & $70.8 \pm 6.8$ & $72.6 \pm 4.1$ \\
Margin & $71.0 \pm 8.0$ & $74.8 \pm 4.5$ & $77.7 \pm 9.6$ \\
Coreset & $71.4 \pm 2.7$ & $75.0 \pm 2.8$ & $76.1 \pm 2.6$ \\
Random & $77.3 \pm 4.1$ & $80.6 \pm 2.4$ & $79.1 \pm 6.9$ \\
DUIS (Ours) & $\mathbf{81.1 \pm 2.0}$ & $\mathbf{84.3 \pm 1.4}$ & $\mathbf{80.0 \pm 2.6}$ \\
\bottomrule
\end{tabular}
\endgroup
\end{table}

Fig.~\ref{fig:real_acc} compares different active learning methods on the real-world single-photon dataset. Our method consistently outperforms the baselines across different labeling budgets. With only 90 labeled samples,  about 8.2\% of the training split, it reaches 90\% accuracy, which is 6\% higher than the best baseline.
Table~\ref{tab:real_cls_msppp} further evaluates the classification performance under different MSPPP regimes. DUIS achieves the best accuracy on all three splits, with $81.1\pm2.0$, $84.3\pm1.4$, and $80.0\pm2.6$ accuracy under high-, medium-, and low-MSPPP conditions, respectively. Compared with the strongest baseline in each regime, DUIS improves accuracy by 3.8, 3.7, and 0.9 percentage points, respectively. In addition, DUIS maintains relatively low variance across the three regimes, suggesting that the proposed condition-aware acquisition strategy improves not only label efficiency but also robustness to photon-density variations.

\begin{table}[!t]
\centering
\caption{Ablation study on the real SPL classification dataset at the fourth active-learning round ($7.27\%$ labeled samples). Syn denotes the common physics-grounded synthetic SPL training used by all variants. Results are reported as mean accuracy $\pm$ standard deviation over repeated runs.}
\label{tab:ablation_duis}
\setlength{\tabcolsep}{4pt}
\renewcommand{\arraystretch}{1.08}
\begingroup
\begin{tabular}{lcccc}
\toprule
Variant & Syn & Diversity & Inconsistency & Accuracy \\
\midrule
Syn + BADGE & \checkmark & -- & --  & 93.4 $\pm$ 4.0 \\
Syn + Entropy & \checkmark & -- & --  &92.0 $\pm$ 2.4 \\
Syn + Margin & \checkmark & -- & --  & 94.8 $\pm$ 4.8\\
Syn + Coreset & \checkmark & -- & -- & 96.2 $\pm$ 1.3 \\
Syn + Random & \checkmark & -- & --  & 97.6 $\pm$ 1.1 \\
Syn + Diversity & \checkmark & \checkmark & --  & 98.8 $\pm$ 1.3 \\
Full (Ours) & \checkmark & \checkmark & \checkmark  & 99.2 $\pm$ 0.5 \\
\bottomrule
\end{tabular}
\endgroup
\end{table}

\subsection{Ablation Study}
\label{sec:exp_ablation}

We conduct an ablation study on the real SPL classification dataset at the fourth active-learning round, corresponding to 80 labeled samples ($7.27\%$). All variants use the same physics-grounded synthetic SPL variants (Syn), initialization, query budget, and training protocol. Table~\ref{tab:ablation_duis} summarizes the results under this controlled setting.

The ablation is interpreted through two paired comparisons. Syn + Random versus Syn + Diversity quantifies the contribution of diversity-based candidate selection, while Syn + Diversity versus Full isolates the additional contribution of photon-condition inconsistency. Since Syn is fixed across all variants, these comparisons isolate the effect of the acquisition rule under a matched synthetic-training setting.

\subsection{Parameter Sensitivity Analysis}
\label{sec:exp_sensitivity}
We assess the sensitivity of DUIS to $N_{\mathrm{cand}}$, $\alpha$, and $M$ by varying one parameter at a time while fixing the others. The $N_{\mathrm{cand}}$ study is performed at the fourth active-learning round, while $\alpha$ and $M$ are evaluated at the fifth round with 100 labeled samples ($9.09\%$). The corresponding results are summarized in Table~\ref{tab:parameter_sensitivity}.

\begin{table}[!t]
\centering
\caption{Parameter sensitivity on the real SPL classification dataset. One parameter is varied at a time while the others are fixed to their default values. $N_{\mathrm{cand}}$ is evaluated at the fourth active-learning round; $\alpha$ and $M$ are evaluated at the fifth round ($9.09\%$ labeled samples). Results are reported as mean accuracy $\pm$ standard deviation over repeated runs.}
\label{tab:parameter_sensitivity}
\begingroup
\scriptsize
\setlength{\tabcolsep}{4pt}
\renewcommand{\arraystretch}{1.08}
\begin{tabular}{@{}ccc@{}}
\toprule
$N_{\mathrm{cand}}$ & $\alpha$ & $M$ \\
\midrule
\begin{tabular}{@{}cc@{}}
Value & Accuracy \\
\midrule
40  & $69.9 \pm 6.9$ \\
80  & $85.7 \pm 6.5$ \\
120 & $92.3 \pm 1.7$ \\
160 & $\mathbf{96.4 \pm 1.8}$ \\
200 & $96.2 \pm 1.3$
\end{tabular}
&
\begin{tabular}{@{}cc@{}}
Value & Accuracy \\
\midrule
0.25 & $90.8 \pm 1.5$ \\
0.50 & $\mathbf{92.4 \pm 2.0}$ \\
0.75 & $90.6 \pm 2.6$
\end{tabular}
&
\begin{tabular}{@{}cc@{}}
Value & Accuracy \\
\midrule
1 & $73.0 \pm 5.3$ \\
2 & $77.7 \pm 2.2$ \\
3 & $76.3 \pm 4.6$ \\
4 & $\mathbf{79.1 \pm 2.6}$
\end{tabular}
\\
\bottomrule
\end{tabular}
\endgroup
\end{table}

TThe candidate-pool size has the strongest effect among the evaluated parameters. Accuracy increases from $69.9 \pm 6.9$ at $N_{\mathrm{cand}}=40$ to $96.4 \pm 1.8$ at $N_{\mathrm{cand}}=160$, with little further improvement at $N_{\mathrm{cand}}=200$, suggesting that a sufficiently large candidate pool is important for effective sample selection.

The results for $\alpha$ are relatively stable, with $\alpha=0.50$ achieving the best mean accuracy of $92.4 \pm 2.0$. For the number of synthetic variants, using multiple conditions consistently outperforms $M=1$, with the best result at $M=4$. Accordingly, we use $\alpha=0.50$ and $M=4$ by default, while selecting $N_{\mathrm{cand}}$ to balance performance and computational cost.

\subsection{Limitations and Future Robotic Extensions}

This study focuses on label-efficient, condition-robust SPL classification and segmentation, rather than a complete robotic system; SPL-based SLAM, semantic mapping, active perception, and closed-loop navigation remain future work.

For robotic deployment, SPL predictions must be associated with robot poses and fused over time, similar to semantic mapping systems that integrate neural predictions into reconstructed maps~\cite{mccormac2017semanticfusion}. The proposed uncertainty-inconsistency score could serve as a reliability cue for such fusion and active data selection, but its integration with SPL-based SLAM and planning remains future work. Practical long-range deployment also requires handling sparse returns, platform motion, limited field of view, calibration, and multi-sensor integration. Although SPL has shown promise for long-range imaging and imaging through atmospheric obscurants~\cite{jiang2023long}, full robotic deployment remains future work.

\section{Conclusion}
\label{section5}

This study addresses the challenge of label-efficient semantic perception from single-photon LiDAR (SPL) data by introducing a condition-aware active learning framework. Rather than treating all SPL samples as being acquired under a fixed sensing condition, the proposed method explicitly uses photon-level imaging variations to assess sample informativeness. By synthesizing SPL variants under different MSPPP and noise conditions, the framework estimates not only prediction uncertainty, but also condition-induced inconsistency, allowing the model to identify samples that are informative and sensitive to realistic sensing variations.

Experiments on SPL image classification and semantic segmentation, including both synthetic and real-world settings, demonstrate that the proposed strategy improves label efficiency and robustness over representative active learning baselines. These results show that acquisition-condition awareness is a useful principle for scalable SPL perception under limited annotation budgets. Future work will extend this framework toward more complex robotic perception scenarios, such as SPL-based semantic mapping, active perception, multi-sensor fusion, and closed-loop navigation.

\section*{ACKNOWLEDGMENT}
The work was supported in part by the National Natural Science Foundation of China
under Grants 62273223, 62336005, 62421004, and 62461160313, and by the
Shanghai Municipal Commission of Education under Grant 24SG38.

\bibliographystyle{IEEEtran}
\bibliography{ref}

@article{cabon2020vkitti2,
  title={Virtual {KITTI} 2},
  author={Cabon, Yohann and Murray, Naila and Humenberger, Martin},
  journal={arXiv preprint arXiv:2001.10773},
  year={2020}
}

@article{kirmani2014first,
  title={First-photon imaging},
  author={Kirmani, Ahmed and Venkatraman, Dheera and Shin, Dongeek and Cola{\c{c}}o, Andrea and Wong, Franco NC and Shapiro, Jeffrey H and Goyal, Vivek K},
  journal={Science},
  volume={343},
  number={6166},
  pages={58--61},
  year={2014},
  publisher={American Association for the Advancement of Science}
}

@article{rapp2017few,
  title={A few photons among many: Unmixing signal and noise for photon-efficient active imaging},
  author={Rapp, Joshua and Goyal, Vivek K},
  journal={IEEE Transactions on Computational Imaging},
  volume={3},
  number={3},
  pages={445--459},
  year={2017},
  publisher={IEEE}
}

@article{li2021single,
  title={Single-photon imaging over 200 km},
  author={Li, Zheng Ping and Ye, Jun Tian and Huang, Xin and Jiang, Peng Yu and Cao, Yuan and Hong, Yu and Yu, Chao and Zhang, Jun and Zhang, Qiang and Peng, Cheng Zhi and others},
  journal={Optica},
  volume={8},
  number={3},
  pages={344--349},
  year={2021},
  publisher={Optical Society of America}
}

@article{jiang2023long,
  title={Long range {3D} imaging through atmospheric obscurants using array-based single-photon LiDAR},
  author={Jiang, Peng-Yu and Li, Zheng-Ping and Ye, Wen-Long and Hong, Yu and Dai, Chen and Huang, Xin and Xi, Shui-Qing and Lu, Jie and Cui, Da-Jian and Cao, Yuan},
  journal={Optics Express},
  volume={31},
  number={10},
  pages={16054--16066},
  year={2023},
  publisher={Optica Publishing Group}
}

@inproceedings{lai2011large,
  title={A large-scale hierarchical multi-view {RGB-D} object dataset},
  author={Lai, Kevin and Bo, Liefeng and Ren, Xiaofeng and Fox, Dieter},
  booktitle={2011 IEEE International Conference on Robotics and Automation},
  pages={1817--1824},
  year={2011},
  organization={IEEE}
}

@article{yao2022dynamic,
  title={Dynamic single-photon {3D} imaging with a sparsity-based neural network},
  author={Yao, Gongxin and Chen, Yiwei and Jiang, Chen and Xuan, Yixin and Hu, Xiaomin and Liu, Yong and Pan, Yu},
  journal={Optics Express},
  volume={30},
  number={21},
  pages={37323--37340},
  year={2022},
  publisher={Optica Publishing Group}
}

@inproceedings{zhai2022scaling,
  title={Scaling vision transformers},
  author={Zhai, Xiaohua and Kolesnikov, Alexander and Houlsby, Neil and Beyer, Lucas},
  booktitle={Proceedings of the IEEE/CVF conference on computer vision and pattern recognition},
  pages={12104--12113},
  year={2022}
}

@article{ren2021survey,
  title={A survey of deep active learning},
  author={Ren, Pengzhen and Xiao, Yun and Chang, Xiaojun and Huang, Po-Yao and Li, Zhihui and Gupta, Brij B and Chen, Xiaojiang and Wang, Xin},
  journal={ACM Computing Surveys (CSUR)},
  volume={54},
  number={9},
  pages={1--40},
  year={2021},
  publisher={ACM New York, NY}
}

@incollection{lewis1994heterogeneous,
  title={Heterogeneous uncertainty sampling for supervised learning},
  author={Lewis, David D and Catlett, Jason},
  booktitle={Machine Learning Proceedings 1994},
  pages={148--156},
  year={1994},
  publisher={Elsevier}
}

@inproceedings{scheffer2001active,
  title={Active hidden markov models for information extraction},
  author={Scheffer, Tobias and Decomain, Christian and Wrobel, Stefan},
  booktitle={International Symposium on Intelligent Data Analysis},
  pages={309--318},
  year={2001},
  organization={Springer}
}

@inproceedings{gal2017deep,
  title={Deep bayesian active learning with image data},
  author={Gal, Yarin and Islam, Riashat and Ghahramani, Zoubin},
  booktitle={International Conference on Machine Learning},
  pages={1183--1192},
  year={2017},
  organization={PMLR}
}

@inproceedings{sener2018active,
  title={Active Learning for Convolutional Neural Networks: A Core-Set Approach},
  author={Sener, Ozan and Savarese, Silvio},
  booktitle={International Conference on Learning Representations},
  year={2018},
  organization={ICLR}
}

@inproceedings{AshZK0A20,
  author    = {Jordan T. Ash and
               Chicheng Zhang and
               Akshay Krishnamurthy and
               John Langford and
               Alekh Agarwal},
  title     = {Deep Batch Active Learning by Diverse, Uncertain Gradient Lower Bounds},
  booktitle = {International Conference on Learning Representations},
  year      = {2020},
  organization={ICLR}
}

@article{muller2023robust,
  title={Robust and efficient depth-based obstacle avoidance for autonomous miniaturized uavs},
  author={M{\"u}ller, Hanna and Niculescu, Vlad and Polonelli, Tommaso and Magno, Michele and Benini, Luca},
  journal={IEEE Transactions on Robotics},
  volume={39},
  number={6},
  pages={4935--4951},
  year={2023},
  publisher={IEEE}
}

@inproceedings{goyal2025robust,
  title={Robust 3d object detection using probabilistic point clouds from single-photon lidars},
  author={Goyal, Bhavya and Gutierrez-Barragan, Felipe and Lin, Wei and Velten, Andreas and Li, Yin and Gupta, Mohit},
  booktitle={Proceedings of the IEEE/CVF International Conference on Computer Vision},
  pages={28417--28427},
  year={2025}
}

@article{unal2023discwise,
  title={Discwise active learning for lidar semantic segmentation},
  author={Unal, Ozan and Dai, Dengxin and Unal, Ali Tamer and Van Gool, Luc},
  journal={IEEE Robotics and Automation Letters},
  volume={8},
  number={11},
  pages={7671--7678},
  year={2023},
  publisher={IEEE}
}

@inproceedings{sheikh2020gradient,
  title={Gradient and log-based active learning for semantic segmentation of crop and weed for agricultural robots},
  author={Sheikh, Rasha and Milioto, Andres and Lottes, Philipp and Stachniss, Cyrill and Bennewitz, Maren and Schultz, Thomas},
  booktitle={2020 IEEE International Conference on Robotics and Automation (ICRA)},
  pages={1350--1356},
  year={2020},
  organization={IEEE}
}

@inproceedings{siddiqui2020viewal,
  title={Viewal: Active learning with viewpoint entropy for semantic segmentation},
  author={Siddiqui, Yawar and Valentin, Julien and Nie{\ss}ner, Matthias},
  booktitle={Proceedings of the IEEE/CVF conference on computer vision and pattern recognition},
  pages={9433--9443},
  year={2020}
}

@inproceedings{gao2020consistency,
  title={Consistency-based semi-supervised active learning: Towards minimizing labeling cost},
  author={Gao, Mingfei and Zhang, Zizhao and Yu, Guo and Ar{\i}k, Sercan {\"O} and Davis, Larry S and Pfister, Tomas},
  booktitle={European Conference on Computer Vision},
  pages={510--526},
  year={2020},
  organization={Springer}
}

@article{kim2021lada,
  title={Lada: Look-ahead data acquisition via augmentation for deep active learning},
  author={Kim, Yoon-Yeong and Song, Kyungwoo and Jang, JoonHo and Moon, Il-Chul},
  journal={Advances in Neural Information Processing Systems},
  volume={34},
  pages={22919--22930},
  year={2021}
}

@inproceedings{suonsivu2024time,
  title={Time-Resolved MNIST Dataset for Single-Photon Recognition},
  author={Suonsivu, Aleksi and Salmela, Lauri and Peretti, Edoardo and Uosukainen, Leevi and Bilcu, Radu Ciprian and Boracchi, Giacomo},
  booktitle={European Conference on Computer Vision},
  pages={127--143},
  year={2024},
  organization={Springer}
}

@article{liu2022photon,
  title={Photon-efficient non-line-of-sight imaging},
  author={Liu, Jianjiang and Zhou, Yijun and Huang, Xin and Li, Zheng-Ping and Xu, Feihu},
  journal={IEEE Transactions on Computational Imaging},
  volume={8},
  pages={639--650},
  year={2022},
  publisher={IEEE}
}

@inproceedings{young2025enhancing,
  title={Enhancing autonomous navigation by imaging hidden objects using single-photon lidar},
  author={Young, Aaron and Batagoda, Nevindu M and Zhang, Harry and Dave, Akshat and Pediredla, Adithya and Negrut, Dan and Raskar, Ramesh},
  booktitle={2025 IEEE International Conference on Robotics and Automation (ICRA)},
  pages={4907--4914},
  year={2025},
  organization={IEEE}
}

@inproceedings{mccormac2017semanticfusion,
  title={Semanticfusion: Dense 3d semantic mapping with convolutional neural networks},
  author={McCormac, John and Handa, Ankur and Davison, Andrew and Leutenegger, Stefan},
  booktitle={2017 IEEE International Conference on Robotics and automation (ICRA)},
  pages={4628--4635},
  year={2017},
  organization={IEEE}
}

@inproceedings{meyer2019importance,
  title={The importance of metric learning for robotic vision: Open set recognition and active learning},
  author={Meyer, Benjamin J and Drummond, Tom},
  booktitle={2019 International Conference on Robotics and Automation (ICRA)},
  pages={2924--2931},
  year={2019},
  organization={IEEE}
}

@article{schmidt2024generalized,
  title={Generalized synchronized active learning for multi-agent-based data selection on mobile robotic systems},
  author={Schmidt, Sebastian and Stappen, Lukas and Schwinn, Leo and G{\"u}nnemann, Stephan},
  journal={IEEE Robotics and Automation Letters},
  year={2024},
  publisher={IEEE}
}

\end{document}